\documentclass[12pt]{article}

\renewcommand{\title}[1]{\begin{center}\bf\Large #1\end{center}}

\renewcommand{\author}[1]{\begin{center}\large #1\end{center}}

\usepackage{latexsym,amsfonts}

\begin{document}

\title{Correlation Functions and Vertex Operators of
  Liouville Theory}

\author{
 George Jorjadze${}^a$
\footnote{email: \tt jorj@rmi.acnet.ge}
 and Gerhard Weigt${}^b$
\footnote{email: \tt weigt@ifh.de} \\
{\small${}^a$Razmadze Mathematical Institute,}\\
  {\small M.Aleksidze 1, 380093, Tbilisi, Georgia}\\
{\small${}^b$Institut f\"ur Physik der Humboldt-Universit\"at zu
  Berlin,}\\
{\small Newtonstra{\ss}e 15, D-12489 Berlin, Germany}}
\begin{abstract}
  We calculate correlation functions for vertex operators with
  negative integer exponentials of a periodic Liouville field, and
  derive the general case by continuing them as distributions.  The
  path-integral based conjectures of Dorn and Otto prove to be
  conditionally valid only. We formulate integral representations for
  the generic vertex operators and indicate structures which are
  related to the Liouville $S$-matrix.

\vspace{0.3cm}

\noindent \hspace{1.7cm} Canonical quantisation; Vertex operators; 
Correlation functions

\end{abstract}

The Liouville theory has fundamentally contributed to the development
of both mathematics \cite{L, P} and physics \cite{Pol}, and beyond it
fascinated with a wide range of applications as a conformal field
theory. Nevertheless, its quantum description is still incomplete. By
canonical quantisation \cite{BCT}-\cite{JW} it could be shown that the
operator Liouville equation and the Poisson structure of the theory,
including the causal non-equal time properties, are consistent with
conformal invariance and locality \cite{OW, JW}, but exact results for
Liouville correlation functions remained rare \cite{W} despite of
ambitious programmes \cite{GS, T}.

In this letter we calculate correlation functions for vertex operators
with generic exponentials of a periodic Liouville field.  The vertex
operators are given in terms of the asymptotic $in$-field of the
Liouville theory \cite{J}, and we formulate for them an integral
representation as an alternative to the formal but still useful
infinite sum of \cite{OW}. However, there is so far no reliable recipe
to use such integral operators directly since the complex powers of
the screening charge operators describing them are not constructed
yet, a problem related to the exact knowledge of the Liouville
$S$-matrix. We calculate therefore first correlation functions for
vertex operators of \cite{OW} with negative integer exponentials and
continue the result analytically as a distribution, as is required by
the zero mode contributions of the Liouville theory \cite{JW1}. We
prove so that the correlation functions suggested in \cite{DO} are
conditionally applicable only. This is indeed a surprise because the
conjecture of \cite{DO} was obtained by standard analytical
continuation of a path-integral result for minimal models \cite{GL},
which describes nothing but a special part of the operator based
correlation function \cite{W}. In this respect it is worth mentioning
that already the Liouville reflection amplitudes \cite{ZZ} proved to
be identical with those obtained from the Liouville $S$-matrix
\cite{JW1, JW2}.

We parametrise the vertex operators by a free-field which allows to
avoid the use of quantum group representations, and we define the
approach on which both, the derivation of the correlation functions
and the formulation of the integral representation of the vertex
operators are based. The structures needed to understand the Liouville
$S$-matrix are indicated, and in the conclusions we stress the
importance of the results for the related WZNW cosets.

\section{Free-field parametrisation}

We use minkowskian light-cone coordinates $x=\tau+\sigma,\, \bar
x=\tau-\sigma$ and select from ref. \cite{L} that general solution of
the Liouville equation
\begin{equation}\label{G-sol}
\varphi = \frac{1}{2}\,\log\,\frac{A'(x)\bar A'(\bar x)}
{[1+\mu^2 A(x)\bar A(\bar x)]^2},
\end{equation}
which has a particularly utilisable physical interpretation. For
periodic boundaries
\begin{equation}\label{P-B}
\varphi(\tau,\sigma+2\pi)=\varphi(\tau,\sigma),
\end{equation}
one can parametrise the non-canonical and quasi-periodic parameter
functions $A(x)$, $\bar A(\bar x)$
\begin{equation}\label{A-mon}
A(x+2\pi)=e^{\gamma p}A(x),~~~~
\bar A(\bar x-2\pi)=e^{-\gamma p}\bar A(\bar x),
\end{equation}
by the canonical free field
$2\phi(\tau,\sigma)=\log\,{A'(x)\bar A'(\bar x)}$ with standard
mode expansion
\begin{equation}\label{FF0}
\phi(\tau,\sigma)=\gamma q+\frac{\gamma p}{2\pi}\tau +
\frac{i\gamma}{\sqrt{4\pi}}\sum_{n\neq 0}\frac{a_n}{n}\, e^{-inx}+
\frac{i\gamma}{\sqrt{4\pi}}\sum_{n\neq 0}\frac{\bar a_n}{n}\, e^{-in\bar x},
\end{equation}
and the chiral decomposition
\begin{equation}\label{FF-ch}
\phi(\tau,\sigma)=\left(\,\frac{1}{2}\gamma q+\frac{\gamma p}{4\pi}x +
\phi(x)\,\right)+\left(\,\frac{1}{2}\gamma q+\frac{\gamma p}{4\pi}\bar x +
\bar\phi(\bar x)\,\right).
\end{equation}
Note the rescalings of the fields $\phi$ and $\varphi$ by the Liouville
coupling $\gamma$. We obtain so the canonical transformation between
the Liouville and the free-field of \cite{OW} ($\gamma$ there is
$2\gamma$ here!)
\begin{equation}\label{FF-par}
e^{-\varphi(\tau,\sigma)}=e^{-\phi(\tau,\sigma)} +
\mu^2 e^{-\phi(\tau,\sigma)} A(x)\bar A(\bar x),
\end{equation}
and by integrating $A'(x)$ ( correspondingly $\bar A'(\bar x)$ ) using
(\ref{A-mon})
\begin{equation}\label{A}
A(x)=\frac{e^{\gamma q}}{2\sinh \frac{\gamma p}{2}}
\int_0^{2\pi}dy\, e^{\frac{\gamma p}{2}\left(\epsilon(x-y)+
\frac{y}{\pi}\right)
+ 2\phi(y)}.
\end{equation}
$\epsilon(z)$ is the stair-step function, and as preconceived, the
non-vanishing parameter $p$ of the hyperbolic monodromy relation
(\ref{A-mon}) becomes identical with the momentum zero mode of the
free field (\ref{FF0}), and we choose $p>0$ \cite{BCT}.

Since for asymptotic 'time' $\tau$ the two terms of the Liouville
exponential (\ref{FF-par}) behave as
\begin{equation}\label{in-out-0-mode}
e^{-\phi(\tau,\sigma)}\sim  e^{-\frac{\gamma p}{2\pi}\tau},~~~~~
\mu^2 e^{-\phi(\tau,\sigma)} A(x)\bar A(\bar x)\sim
e^{\frac{\gamma p}{2\pi}\tau},
\end{equation}
they can be interpreted as $in$-coming respectively $out$-going
contributions so that
\begin{equation}\label{Liouville-in-out}
e^{-\varphi(\tau,\sigma)}=e^{-\phi_{in}(\tau,\sigma)}+
e^{-\phi_{out}(\tau,\sigma)}.
\end{equation}
As a consequence the chosen canonical free field (\ref{FF0}) has a 
physical meaning, it is the asymptotic $in$-field of the Liouville 
theory \cite{J}, and the $out$-field which is given by the $in$-field 
too
\begin{equation}\label{out-field}
e^{-\phi_{out}(\tau,\sigma)}=
\mu^2 e^{-\phi(\tau,\sigma)}\,A(x)\,\bar A(\bar x),
\end{equation}
defines likewise the classical form of the Liouville $S$-matrix.

Two related forms of the Liouville exponential are relevant 
for canonical quantisation, the formal expansion in powers of
the conformal weight zero functions $A(x)\bar A(\bar x)$ \cite{OW}
\begin{equation}\label{V_l}
e^{2\lambda\varphi(\tau,\sigma)}=e^{2\lambda\phi(\tau,\sigma)}
\sum_{m= 0}^\infty\frac{(-1)^m}{m!} \frac{\Gamma(2\lambda+m)}
{\Gamma(2\lambda)}[\mu^2A(x)\bar A(\bar x)]^m,
\end{equation}
and the integral representation for positive $\lambda$
\begin{equation}\label{V_l-I}
e^{2\lambda\varphi(\tau,\sigma)}=
e^{2\lambda\phi(\tau,\sigma)}
\int_{-\infty}^{+\infty}dk\,\,\frac{\Gamma\left(\lambda+ik\right)\,
\Gamma\left(\lambda-ik\right)}{2\pi\Gamma(2\lambda)}\,
[\mu^2A(x)\bar A(\bar x)]^{-(\lambda+ik)}.
\end{equation}
The last equation follows from the Liouville solution (\ref{FF-par}) 
by using the Fourier transformation of $(2\cosh y)^{-2\lambda}$ \cite{GR}
\begin{equation}\label{A(k)}
\int_{-\infty}^{+\infty}dk\,\,\frac{\Gamma\left(\lambda+ik\right)\,
\Gamma\left(\lambda-ik\right)}{2\pi\Gamma(2\lambda)}\,\, e^{2iky}
=\left(e^{-y}+e^y\right)^{-2\lambda}
 ~~~~~\mbox{for}~~~~~\mbox{Re}\,\lambda>0.
\end{equation}
If we continue this equation from positive to negative $\lambda$ 
and consider the kernel of that integral as a generalised function
( see also eqs. (56) - (61) of \cite{JW1} ), for $\lambda\rightarrow 
-\,n/2$ we obtain
\begin{equation}\label{A_n}
\frac{\Gamma\left({\lambda+ik}\right)\,
\Gamma\left({\lambda-ik}\right)}{2\pi\Gamma(2\lambda)}~~\longrightarrow~~
\sum_{m=0}^n{n \choose m}\,\delta(k-i(n/2-m)).
\end{equation}
In this manner (\ref{V_l-I}) becomes in fact identical with the 
corresponding finite sum of (\ref{V_l}).

It might be interesting to notice here that the investigations
initiated by the references \cite{BCT, Neveu} are based on Liouville's
second form of the general solution \cite{L}. This approach led
\cite{BCT} to a parametrisation of the Liouville theory in terms of a
canonically related \cite{FS} but pseudo-scalar free field which is
asymptotically neither an $in$- nor an $out$-field, whereas the work
of \cite{Neveu, GS} mainly treats the singular elliptic monodromy for
which we do not know whether there exists a parametrisation in terms
of a real free field at all.

\section{Vertex operators}

The quantum Liouville theory will be defined by canonically
quantising the free field (\ref{FF0})
\begin{equation}\label{CC}
[q, p]=i \hbar,~~~~~~[a_m, a_n]=\,\hbar\,m\,\delta_{m+n,0},
\end{equation}
and requiring that the vertex operators are primary and local.  Such a
procedure gives an anomaly-free quantum Liouville theory only if
additional quantum deformations are taken into consideration
\cite{BCT} - \cite{JW}. A quantum realisation of (\ref{V_l}),
consistent with the operator Liouville equation and the canonical
commutation relations, was so constructed in \cite{OW}. But the
infinite sum is not a useful vertex operator. It is its finite form
for negative integer $2\lambda=-n$ which presents a well-defined basis
for the calculation of correlation functions.

Let us review the needed elements, and call
$V_{2\lambda}(\tau,\sigma)$ the vertex operator of 
$e^{2\lambda\varphi(\tau,\sigma)}$.
Taking, for simplicity, the same notations for the classical and
the corresponding normal ordered quantum expressions, the vertex
operator for $\lambda=-1/2$ can be written as
\begin{equation}\label{V}
V_{-1}(\tau,\sigma)=e^{-\phi(\tau,\sigma)}+\mu_\alpha^2\,
\frac{1}{2\sinh \pi P}\,
e^{-\phi(\tau,\sigma)}\,S(\tau,\sigma)\,\frac{1}{2\sinh \pi P}.
\end{equation}
Here $\mu_\alpha^2$ is the renormalised 'cosmological constant'
and we introduced short notations
\begin{equation}\label{mu}
\mu^2_\alpha =\mu^2\,\frac{\sin\pi\alpha}{\pi\alpha}, ~~~~~
\alpha=\frac{\hbar\gamma^2}{2\pi},~~~~~P=\frac{\gamma p}{2\pi},
\end{equation}
and defined  the conformal weight zero screening charge operator as
\begin{equation}\label{S(x)}
S(\tau,\sigma)=e^{\gamma q+P\tau} A_p(x)\bar A_p(\bar x)e^{\gamma
q+P\tau},
\end{equation}
where
\begin{equation}\label{A_p}
A_p(x)=\int_0^{2\pi}dy\, e^{2\phi(x+y)+P(y-\pi)}
\end{equation}
is the integral (\ref{A}) rewritten by using the periodicity and
$\epsilon(z)=sign(z)$ for $z\in(-2\pi,2\pi)$.

The useful factorised form of the operator $V_{-n}$ can be 
constructed easily by induction
\begin{equation}\label{V_n+1}
V_{-n-1}(x,\bar x)=\lim_{\epsilon\rightarrow 0} V_{-n}(x,\bar
x)\,V_{-1}(x+\epsilon,\bar x-\epsilon)\, \epsilon^{\alpha\,n},
\end{equation}
where the regularising factor $\epsilon^{\alpha\,n}$ just removes the 
short distance singularity, and it results
\begin{equation}\label{V_n}
V_{-n}(\tau,\sigma)=\sum_{m=0}^n C_m^n\,\mu_\alpha^{2m}\,
\prod_{l=1}^m\frac{1}{2\sinh \pi(P+i\, l\alpha)} \,
e^{-n\phi(\tau,\sigma)}\,S^m(\tau,\sigma)\,
\prod_{l=1}^m\frac{1}{2\sinh \pi(P-i\,l\alpha)}.
\end{equation}
The shift of the momenta is a consequence of locality, and the
same holds for the deformed binomial coefficients
\begin{equation}\label{C_n^m}
C_m^n=\prod_{l=1}^m\frac{\sin\pi(n-l+1)\alpha}{\sin\pi l\alpha},
\end{equation}
whereas the hidden short distance contributions of (\ref{V_n})
are due to the conformal properties.

Note that $[e^{-\phi(\tau,\sigma)}, S(\tau,\sigma)]=0$ provides 
hermiticity of (\ref{V}). The vertex operators for arbitrary 
$\lambda$ will be described jointly with the correlation functions 
in the next section.

\section{Correlation functions}

Owning to conformal invariance we have to calculate 3-point
correlation functions only. They are defined by matrix elements of
vertex operators $\langle p; 0|\,V_{2\lambda}(0,0)\, |p'; 0\rangle$
between the highest weight vacuum state $|p; 0\rangle$ ($~p>0~$) which
gets annihilated by the operators $a_n$ with $n>0$.  Using (\ref{V_n})
the correlation function for $2\lambda=-n$ becomes ( $P=\frac{\gamma
p}{2\pi}$ ! )
\begin{eqnarray}\label{V_pp}
\langle p; 0|\,V_{-n}(0,0)\, |p'; 0\rangle = \sum_{m=0}^n
C_m^n\,\mu_\alpha^{2m}\, \,J^n_m(P,\alpha)\,
\left(I^n_m(P,\alpha)\right)^2\,\delta(P-P\,'-i(n-2m)\alpha),
\end{eqnarray}
where $J^n_m(P,\alpha)$ summarises the $p$-dependent factors of
the $\sinh$-terms of (\ref{V_n})
 \begin{equation}\label{J_n^m}
J^n_m(P,\alpha)= \prod_{l=1}^m
\frac{1}{4\,\sinh\pi(P+il\alpha)\,\, \sinh\pi(P-i(n-2m+l)\alpha)},
\end{equation}
and $I^n_m(P,\alpha)$ is the (anti-)chiral matrix element
$\langle\, 0|\, e^{-n\phi }\, \prod_{l=1}^m
A_{P-i(n-2l+1)\alpha}\,|0\,\rangle$ which is given by the
integrals over the conformal short-distance deformations of
(\ref{V_n})
\begin{eqnarray}\label{I'_n^m}
I^n_m(P,\alpha)=\int_0^{2\pi}dy_1\,...\,\int_0^{2\pi}dy_m\,
\prod_{l=1}^m e^{(P-\frac{i}{2}n\alpha+im\alpha)(y_l-\pi)}\,
\left(2\sin\frac{y_l}{2}\right)^{n\alpha} \,\times\\ \nonumber
\prod_{1\leq k<l\leq
m}\left(4\sin^2\frac{y_k-y_l}{2}\right)^{-\alpha}\,
e^{i\pi\epsilon(y_k-y_l)}.
\end{eqnarray}
Fortunately these integrals can be expressed \cite{GS} by
Dotsenko-Fateev integrals \cite{FD}
\begin{eqnarray}\label{I''_n^m}
I^n_m(P,\alpha)=
\prod_{l=1}^m\frac{\Gamma(1+(n-l+1)\alpha)}{\Gamma(1+l\alpha)}\,\,
\frac{2\pi\,\,\Gamma(1+\alpha)}
{\Gamma(1+l\alpha-iP)\Gamma(1+(n-2m+l)\alpha+iP)}.
\end{eqnarray}
If we replace the $\sin$- ($\sinh$-) functions of (\ref{C_n^m}) 
( respectively (\ref{J_n^m}) ) by $\Gamma$-functions
\begin{equation}\label{sin-Gamma}
\frac{\pi x}{\sin\pi x}=\Gamma(1+x)\,\Gamma(1-x),
\end{equation}
 we find for the correlation function the result
\begin{eqnarray}\label{V_pp'}
\langle p; 0|\,V_{-n}(0,0)\, |p'; 0\rangle=
\sum_{m=0}^n {n \choose m}\left(\mu^2\frac{\Gamma(1+\alpha)}
{\Gamma(1-\alpha)}\right)^{m} V_m^n(P,P')
\,\delta(P-P'-i(n-2m)\alpha),
\end{eqnarray}
with
\begin{eqnarray}\label{V_m^n}
V_m^n(P,P\,')=\prod_{l=1}^m\frac{~~~~~~\Gamma(iP-l\alpha)\,\,
\Gamma(-iP\,'-l\alpha)\,\,\Gamma(1+(n-l+1)\,\alpha)\Gamma(1-l\alpha)\,
}{\Gamma(1-iP+l\alpha)\,
\Gamma(1+iP\,'+l\alpha)\,\Gamma(1+(l-n-1)\,\alpha)\,\Gamma(1+l\alpha)
}\,.
\end{eqnarray}
Our aim is to continue these functions from the negative value 
$2\lambda=-n$ to positive $\lambda$. We apply eq. (\ref{A_n}) and
obtain, with 
$2\alpha k=P-P'$ from the continued $\delta$-function of (\ref{V_pp'}),
\begin{equation}\label{V_2}
\langle p; 0|\,V_{2\lambda}(0,0)\, |p'; 0\rangle
=\frac{\Gamma\left({\lambda+ik}\right)\,
\Gamma\left({\lambda-ik}\right)}{4\pi\alpha\,\Gamma(2\lambda)}\,
\left(\mu^2\,\,\frac{\Gamma(1+\alpha)}
{\Gamma(1-\alpha)}\right)^{ik-\lambda}\,V_{ik-\lambda}^{-2\lambda}(P,P\,'),
\end{equation}
where $V_{ik-\lambda}^{-2\lambda}(P,P\,')$ is the analytical
continuation of (\ref{V_m^n}). This continuation will be performed by 
means of the integral representation of the $\Gamma$-function \cite{GR}
\begin{equation}\label{log-Gamma}
\log \Gamma(x)=\int_0^\infty \frac{dt}{t}\left[
\frac{e^{-xt}-e^{-t}}{1-e^{-t}} +(x-1)e^{-t}\right],
\end{equation}
and the following summation under that integral
\begin{eqnarray}\label{f}
f(x,\,\alpha|\,m)= \sum_{l=0}^{m-1}\log\Gamma(x+l\alpha)=
~~~~~~~~~~~~~~~~~~~~~~~~~~~~~~~~~~
\\ \nonumber
\int_0^\infty \frac{dt}{t}\left[\frac{e^{-xt}\,(1-e^{-m\alpha t})}
{(1-e^{-t})(1-e^{-\alpha t})}-
\frac{me^{-t}}{1-e^{-t}}+m(x-1)e^{-t}+\alpha
\frac{m(m-1)}{2}\,e^{-t}\right].
\end{eqnarray}
The function $f(x,\,\alpha|\,m)$ has a natural analytical continuation 
with respect to $\alpha$, $m$ and $x$, and the useful property
$f(x-m\alpha,\,\alpha|\,m)=f(x-\alpha,\,-\alpha|\,m)$. To simplify our 
result we rewrite a factor of (\ref{V_m^n})
\begin{equation}\label{G/G}
\prod_{l=1}^m\frac{1}{\Gamma(1+(l-n-1)\,\alpha)}
=\frac{1}{\alpha^m}\,\prod_{l=1}^m\frac{1}{l-n-1}\,
\prod_{l=1}^m\frac{1}{\Gamma((l-n-1)\,\alpha)},
\end{equation}
and continue it separately
\begin{equation}\label{G-G1}
\frac{1}{\alpha^{ik-\lambda}}\,\frac{\Gamma(2\lambda)}{\Gamma(ik+\lambda)}\,
e^{-f(2\lambda\alpha,\,\alpha|\,m)}.
\end{equation}
After analytically continuing the remaining terms of (\ref{V_m^n}),
and obvious cancellations, we obtain finally the generic correlation 
function we are looking for as
\begin{equation}\label{V_3}
\langle p; 0|\,V_{2\lambda}(0,0)\, |p'; 0\rangle
=\frac{\Gamma(\lambda-ik)}{4\pi\alpha}\,
\left(\frac{\mu^2}{\alpha}\,\,\frac{\Gamma(1+\alpha)}
{\Gamma(1-\alpha)}\right)^{ik-\lambda}\,\prod_{j=1}^4
e^{f(x_j,\,\alpha|\,m)-f(y_j,\,\alpha|\,m)},
\end{equation}
where
\begin{eqnarray}\label{x-j,y_j}
&&x_1=iP-m\alpha, ~~~~~x_2=-iP'-m\alpha, ~~~~~ x_3=
1+(n-m+1)\alpha,~~~ x_4=1-m\alpha;
\nonumber\\
&&y_1=1+\alpha-iP, ~~~ y_2=1+\alpha+iP', ~~~~~ y_3=
-n\alpha,~~~~~~~~~~~~~~~~~~~~y_4=1+\alpha;
\nonumber\\
&&m=ik-\lambda, ~~~~~~~~~~n=-2\lambda,~~~~~~~k=\frac{P-P'}{2\alpha},~~~~~~
\alpha=\frac{\hbar\gamma^2}{2\pi},~~~~~~P=\frac{\gamma p}{2\pi}.
\end{eqnarray}

It is worth mentioning here that the function (\ref{f}) was used for
the parametrisation of the 3-point correlation functions suggested in
\cite{DO}, and it is easy to show that with eqs. (\ref{V_3})-
(\ref{x-j,y_j}) we have re-derived that result (see eq. (14) of the
second reference of \cite{DO} with obvious changes of the notation and
overall renormalization).

However, there are some further remarks in order. Dorn and Otto
\cite{DO} started their analytical continuation from a path-integral
result for minimal models \cite{GL} which is just the one term $n=2m$
of the operator calculated correlation function (\ref{V_pp'})
proportional to $\delta(P-P')$. This single term would be selected in
our calculations if and only if screening charge conservation could be
operative \cite{W}. However, the Liouville theory is M\"obius
non-invariant and all the $(n+1)$ terms of (\ref{V_pp'}) together
characterise this theory for $2\lambda=-n$. Moreover, for odd $n$ the
correlation functions of \cite{DO} vanish and only the neglected $n$
terms guarantee the necessary non-vanishing of the Liouville
correlation functions in those points. Vice versa, by analytically
continuing the correlation function of \cite{DO} as generalised
function, in the manner explicitely described for the zero modes in
ref. \cite{JW1}, one finds the correlation function for negative
$\lambda$ which for $\lambda=-n/2$ just reproduces (\ref{V_pp'}).

We should, furthermore, mention that the functions (\ref{f}) and 
$\Upsilon_b$ of \cite{ZZ} are related by
\begin{eqnarray}\label{Upsilon}
f(bu,\,b^2\,|s)-
f(bv,\,b^2\,|s)=\log\Upsilon_b(v)-\log\Upsilon_b(u)+ sb(u-v)\log b,
\nonumber \\
f(1-sb^2,\,b^2\,|s)- f(1+b^2,\,b^2\,|s)=
\log\frac{\Upsilon_b(1/b)}{\Gamma(-s)\,\Upsilon_b(-sb)}+ (s+1)(1-sb^2)\log b,
\end{eqnarray}
where \,$b^2=\alpha$, and $u+v=Q-sb$ with  $Q=b+1/b$. With these 
equations we can derive from (\ref{V_3}) the more heuristically 
motivated alternative, but to \cite{DO} equivalent, correlation 
functions of \cite{ZZ} as follows
\begin{equation}\label{Upsilon3}
4\pi\alpha\langle p; 0|\,V_{2\lambda}(0,0)\, |p'; 0\rangle=
C(\, \frac{1}{2}\left(Q-i{P}/{b}\right),\,\,b\lambda,\,\,
\frac{1}{2}\left(Q+i\,{P\,'}/{b}\right)\, ).
\end{equation}

By the same procedure of analytical continuation as used before, and 
by taking into consideration the results of this section, we obtain
from (\ref{V_n}) the vertex operator for positive $\lambda$ as an  
integral representation
\begin{eqnarray}\label{V_lambda}
V_{\lambda}(\tau,\sigma)= \int_{-\infty}^{+\infty}dk\,\,
\frac{\Gamma(\lambda+ik)\,\Gamma(\lambda-ik)}
{2\pi\Gamma(2\lambda)}\,
\left(\frac{\mu^2}{2\pi\alpha\,\Gamma(1+\alpha)
\Gamma(1-\alpha)}\right)^{ik-\lambda}\,\times \\ \nonumber
\\ \nonumber
Y_\alpha(\lambda,\,k)\,
X_a(p,\,ik-\lambda)\,
e^{2\lambda\phi(\tau,\sigma)}\,S^{ik-\lambda}(\tau,\sigma)
\,X_a^*(p,\,ik-\lambda),
\end{eqnarray}
with
\begin{eqnarray}\label{Y_a}
Y_\alpha(\lambda,k)=\frac{e^{f(1+\alpha,\, \alpha |ik-\lambda)+
f(1-(ik-\lambda)\alpha,\, \alpha |ik-\lambda)}}
{e^{f(1+2\alpha\lambda,\, \alpha |ik-\lambda)+ f(1-(\lambda
+ik-1)\alpha,\, \alpha |ik-\lambda)}},
\end{eqnarray}
\begin{eqnarray}\label{X_a}
X_a(p,\lambda)=e^{f(1+iP-(ik-\lambda)\alpha,\, \alpha
|ik-\lambda)+ f(1-iP+\alpha,\, \alpha |ik-\lambda)}
\,\frac{\Gamma(1+ik-\lambda -iP/\alpha)}{\Gamma(1-iP/\alpha)}.
\end{eqnarray}

But we should emphasize here that we do not have so far a recipe at
hand to calculate the Liouville correlation functions for positive
$\lambda$ directly from this integral. It would require the knowledge
of complex powers of the screening charge operator
$S^{ik-\lambda}(\tau,\sigma)$, which incidentally would give the
$S$-matrix in compact form too. But this problem is at present under
investigation only, and that is the reason why we cannot compare our
vertex operator (\ref{V_lambda}) with the Ansatz of \cite{T} for which
corresponding screening charge operators are not given either, and as
well no recipe how to treat that vertex for an operator calculation of
correlation functions directly.

\section{Conclusions}

With a suitable free-field parametrisation ad hand, canonical
quantisation proves to be a straightforward and reliable approach for
a description of the quantum Liouville theory. We have calculated the
correlation functions for generic vertex operators by using known
algebraic quantum structures of the theory and their distributional
properties. But the derived integral vertex operators could not be
applied directly since the complex powers of screening charge
operators are not yet constructed. The exact $S$-matrix is therefore
not available too. Nevertheless, it is known that self-adjointness of
the Liouville theory as well as the reflection amplitudes follow from
the $S$-matrix \cite{JW2}, which can be derived at least level by
level. We can so conclude that we have got, in principle, a complete
understanding of quantum Liouville theory. The remaining problems to
be solved are mostly of technical nature.

Since the Liouville theory and the $SL(2;R)/U(1)$ respectively
$SL(2;R)/R_+$ black hole cosets can be derived from the same $SL(2;R)$
WZNW theory by Hamiltonian reduction \cite{JW3}, we expect also a
joint quantum treatment of them. While doing so, the Liouville theory
is an important ingredient of the other cosets, so that its quantum
description is a prerequisite for the quantisation of the other
cosets. This might be relevant for $AdS_3$ and string theory too, and
different boundary conditions should be taken into consideration.

Moreover, we believe that the observed causal non-equal time
structures of the cosets are important for field theory in general,
and that these two-dimensional conformal field theories will remain
outstanding examples of mathematical physics even in the next future.

\vspace{0.5cm}
\noindent
{\bf {\Large Acknowledgements}}

\noindent We are especially grateful to Harald Dorn for explaining us
in detail the calculations of ref. \cite{DO}, and thank Chris Ford and
Martin Reuter for discussions.  G.J. is grateful to DESY Zeuthen and
ICTP for hospitality. The research was supported by grants from the
DFG, INTAS, RFBR and GAS.

\end{document}